\documentclass[twocolumn,twoside,slac]{revtex4}
\usepackage{graphicx}
\usepackage{fancyhdr}
\begin{document}

\title{The Geant4 Hadronic Verification Suite for the Cascade Energy Range}

\author{V. Ivanchenko}
\affiliation{BINP, Novosibirsk, 630090, Russia}
\affiliation{CERN, Geneve, CH 1211, Switzerland}
\author{G. Folger, J.P. Wellisch}
\affiliation{CERN, Geneve, CH 1211, Switzerland}
\author{T. Koi, D.H. Wright}
\affiliation{SLAC, Stanford, CA 94025, USA}

\begin{abstract}
A Geant4 hadronic process verification suite has been designed to test
and optimize Geant4 hadronic models in the cascade energy range.  It
focuses on quantities relevant to the LHC radiation environment and
spallation source targets.  The general structure of the suite is presented,
including the user interface, stages of verification, management
of experimental data, event generation, and comparison of results to data.
Verification results for the newly released Binary cascade and Bertini
cascade models are presented.

\end{abstract}

\maketitle

\thispagestyle{fancy}

\section{INTRODUCTION}
The Geant4 toolkit \cite{G4} includes a collection of models and packages for
hadronic physics which are applicable to various particle transport problems
\cite{HadStatus}.
In Geant4, final state generation is separated from the access and use of cross
sections and from tracking.  It is therefore possible and desirable to
have independent, alternative physics models.  In order to optimize these
models and to identify their most appropriate application, detailed
comparisons between experimental data and model predictions are required.

The cascade energy region from the reaction threshold to incident hadron
energies of a few GeV is problematic for all existing Monte Carlo packages.
In order to verify and further develop Geant4 models in this energy range, a
verification suite has been created.  The stages of the verification,
including experimental data handling, event generation and comparison to data
are described here.  Results of the comparison are also presented.

\section{VERIFICATION SUITE}

\subsection{Method}

The verification suite is generic, as it is based on an abstract interface to
a final state hadronic interaction generator.  This interface can be found in
Level 2 of the Geant4 hadronic physics framework \cite{G4had}.  It focuses on
quantities relevant to the LHC radiation environment and spallation source targets.

The modular structure of Geant4 allows the generation of single events with a known
incident particle energy and an explicitly defined hadronic final state generator.
The kinematics of secondaries produced in the interaction are analyzed and the
resulting angular, momentum, energy, and baryon number spectra are stored in
histograms.  The energy-momentum balance can be controlled as well.  The histograms
are compared to published measurements of the differential and double differential
cross sections, $d\sigma/dE$, $d\sigma/d\Omega$, $d^2\sigma/dEd\Omega$, and the
invariant cross sections, $Ed^3\sigma/d^3p$.

The cross section contributing to the $i$-th bin of the histogram are given by
\begin{equation}
\Delta \sigma_i = \sigma_{tot} N_i/N,
\end{equation}
where $\sigma_{tot}$ is the total cross section for the interaction being tested,
$N$ is the number of simulated events in the sample, and $N_i$ is the number of
times bin $i$ is incremented during the run.  Each bin represents a small region of
phase space such as $\Delta \Omega$, $\Delta E$, or $\Delta \Omega \Delta E$, into which
the secondary particle goes after being produced in the interaction.
The double differention cross section is estimatated as
\begin{equation}
\frac{d^2\sigma}{dE d\Omega} = \frac{\Delta  \sigma_i}{\Delta E \Delta \Omega}.
\end{equation}
The verification suite is organized into a number of test cases, each defined by a
unique incident beam energy and a single target nucleus.  All data files, macro files,
kumac files and results for each case are stored in a separate subdirectory.
The AIDA abstract interface \cite{AIDA} is used in the suite for histogram
creation and  filling. The output
is done in ASCII and HBOOK formats.

\subsection{Cross Sections}\label{data}

The verification is done by comparing simulation results with experimental data mainly
from the EXFOR database \cite{EXFOR}.  Only data with absolute measurements of the
differential cross sections are utilized in the suite.  The data downloaded from
the database are re-formatted in order to provide intermediate files acceptable
for PAW analysis.  In some cases re-binning of the data was performed during this process.
This was required mainly for the low energy part of the spectra.  If, within a given
test case, only double differential cross section data are available, the single
differential cross sections are obtained by numerical integration.

Initial verifications have been performed for neutrons and pions produced by protons
incident upon various targets.  For these test cases the initial proton energies were all
below 1 GeV.  In this energy region the inclusive reaction channel
\begin{equation}
p + A \rightarrow n + X,
\end{equation}
has been studied experimentally for many years.  The secondary neutrons can be
identified and their energies can be measured with good precision by using time-of-light
techniques.  Because this reaction is important for many applications, a significant number
of test cases have been created for it (Table~\ref{t_pn}).

\begin{table}[t]
\begin{center}
\caption{Neutron production by incident protons.}
\begin{tabular}{|c|c|}
\hline
\textbf{Target nucleus} & \textbf{Beam energy(MeV)}\\
\hline Be & 113, 256, 585, 800 \\
\hline C & 113, 256, 590 \\
\hline Al & 22, 39, 90, 113, 160, 256, 585, 800 \\
\hline Fe & 22, 65, 113, 256, 597, 800 \\
\hline Ni & 585\\
\hline Zr & 22, 35, 50, 90, 120, 160, 256, 800 \\
\hline Pb & 35, 65, 113, 120, 160, 256, 597, 800 \\
\hline
\end{tabular}
\label{t_pn}
\end{center}
\end{table}

For HEP applications and in particular for the LHC detector, simulating secondary
pion production is important.  Model verification for pion production by incident protons
is available for several test cases (Table~\ref{t_ppi}) for the reactions
\begin{equation}
p + A  \rightarrow \pi^{\pm} + X.
\end{equation}

\begin{table}[t]
\begin{center}
\caption{Pion production by incident protons.}
\begin{tabular}{|c|c|}
\hline
\textbf{Target nucleus} & \textbf{Beam energy(MeV)}\\
\hline H & 585 \\
\hline D & 585 \\
\hline Be & 585 \\
\hline C & 590 \\
\hline Al & 585, 730, 1000 \\
\hline Cu & 730 \\
\hline Ni & 200, 585\\
\hline Pb & 585, 730 \\
\hline
\end{tabular}
\label{t_ppi}
\end{center}
\end{table}

\subsection{User Interface}
The user interface is implemented by macro files which allow the specification
of various parameters of the verification.  These include:
\begin{itemize}
\item initial particle and its energy,
\item beam energy spread,
\item target nucleus,
\item hadronic interaction generator,
\item generator options, and
\item histogram types and bins.
\end{itemize}
Both linear and logarithmic binning are available.  Each verification test case
contains a script which executes the macro files.  Hence the user can perform
the complete verification process by issuing a single command.  The results are
stored in HBOOK format and can be processed in PAW by prepared kumac files.




\section{HADRONIC MODELS}
The verification suite has been used extensively during the release phase
of the new Geant4 Bertini cascade  and Binary cascade
packages \cite{HadStatus}.  The Bertini cascade model is a classical cascade code which was
described in detail at another presentation at this conference \cite{Bertini}.

The Binary cascade
introduces a new approach to cascade calculations.  The interaction
is modeled exclusively on binary scattering between reaction participants and nucleons. 
The nucleus is described by a detailed 3-dimensional model in which the nucleons are
explicitly positioned in phase space.  Free hadron-hadron elastic and reaction
cross sections are used to define the collisions.  Propagation of the particles in the
nuclear field is done by numerically solving the the equation of motion.

The cascade begins with a projectile and a description of the nucleus,
and terminates when the both average and maximum energy of all particles
within the nuclear boundary are below a given threshold.  The remaining nuclear fragment
is treated by pre-equilibrium decay and de-excitation models \cite{deexc}.

\section{RESULTS OF VERIFICATION}

\subsection{Inclusive neutron spectra}

\begin{figure}
\includegraphics{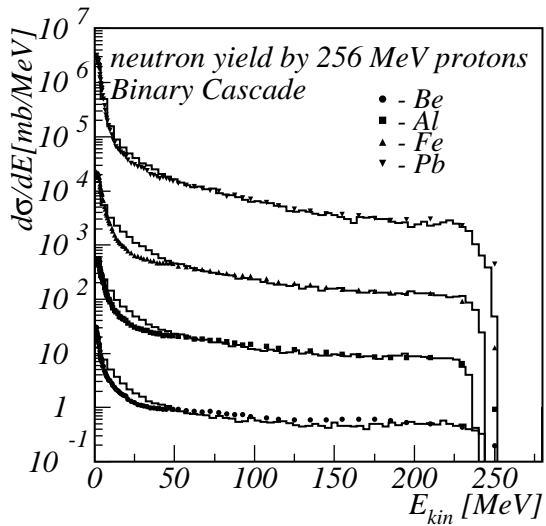}%
\caption{Neutron yield produced  by 256 MeV protons. Histograms - Binary
Cascade predictions, points - data \cite{Meier92}. \label{fig1}}
\end{figure}

\begin{figure}
\includegraphics{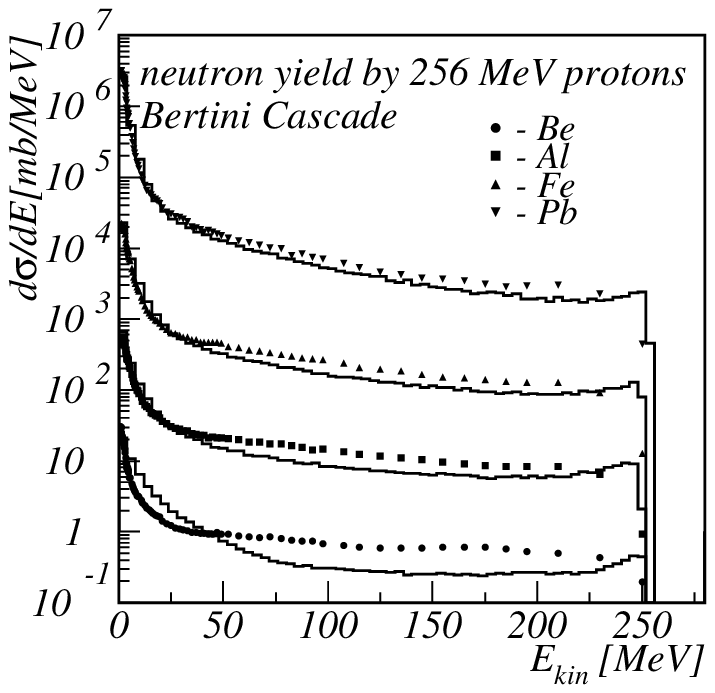}%
\caption{Neutron yield produced  by 256 MeV protons. Histograms - Bertini
Cascade predictions, points - data \cite{Meier92}. \label{fig2}}
\end{figure}

The single and the double differential inclusive neutron spectra
are very sensitive to the physics model used in the cascade code.
As shown in Fig. \ref{fig1}, the Binary cascade reproduces the data 
rather well for all targets, except for energies below 50 MeV
where neutron evaporation is important.
In contrast, the Bertini cascade (Fig.\ref{fig2}) does well below 
50 MeV for all but the lightest nuclei. Above 50 MeV only the general 
trend of the data is reproduced, while the overall normalization improves
as $A$ increases.

\begin{figure}
\includegraphics{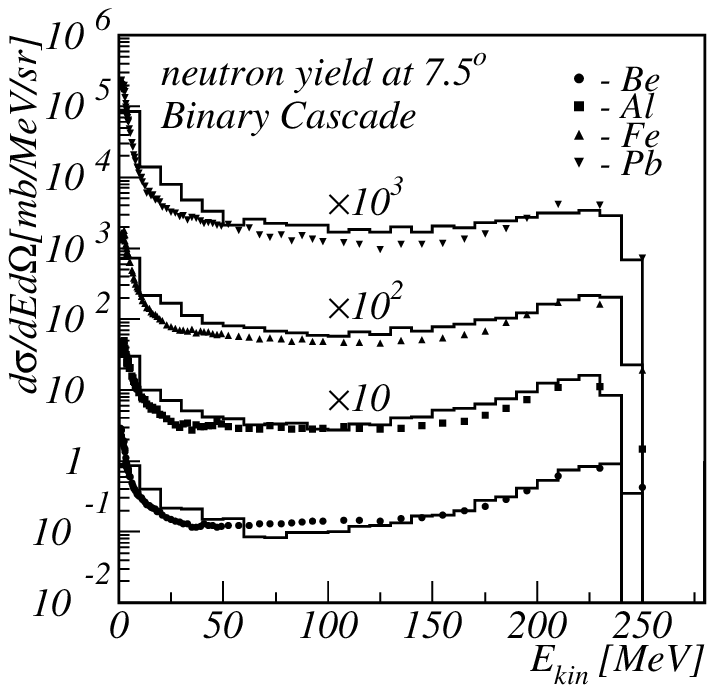}%
\caption{Double differential cross-section for neutrons
produced at 7.5 degrees   by 256 MeV protons. Histograms - Binary
Cascade predictions, points - data \cite{Meier92}. \label{fig3}}
\end{figure}

\begin{figure}
\includegraphics{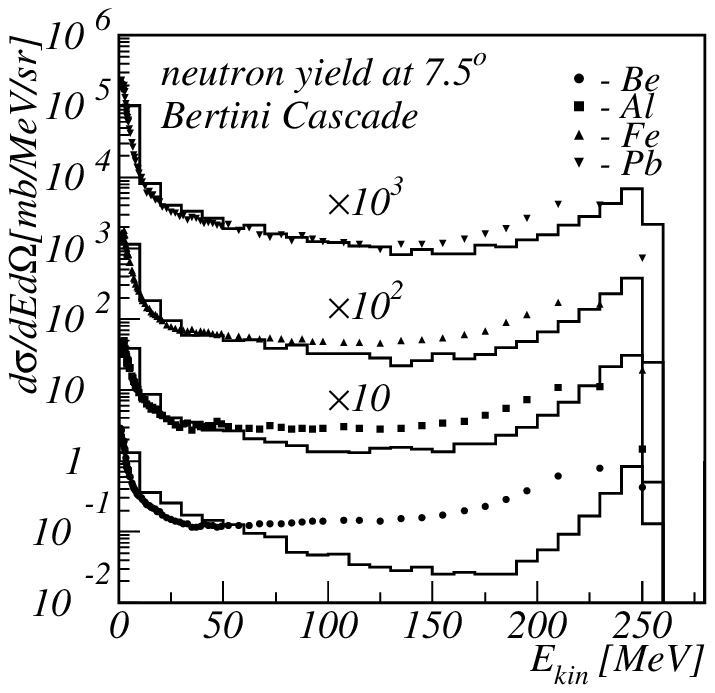}%
\caption{Double differential cross-section for neutrons
produced at 7.5 degrees   by 256 MeV protons. Histograms - Bertini
Cascade predictions, points - data \cite{Meier92}. \label{fig4}}
\end{figure}

Fig. \ref{fig3} shows forward scattering in which the proton transfers its energy
to one target neutron.  Here the Binary cascade describes the data above 50 MeV.
The Bertini cascade (Figs.\ref{fig4}) does less well, significantly 
underestimating the data at higher neutron energies.  This discrepancy is reduced 
for heavier nuclei.

\begin{figure*}
\includegraphics[width=120mm,height=100mm]{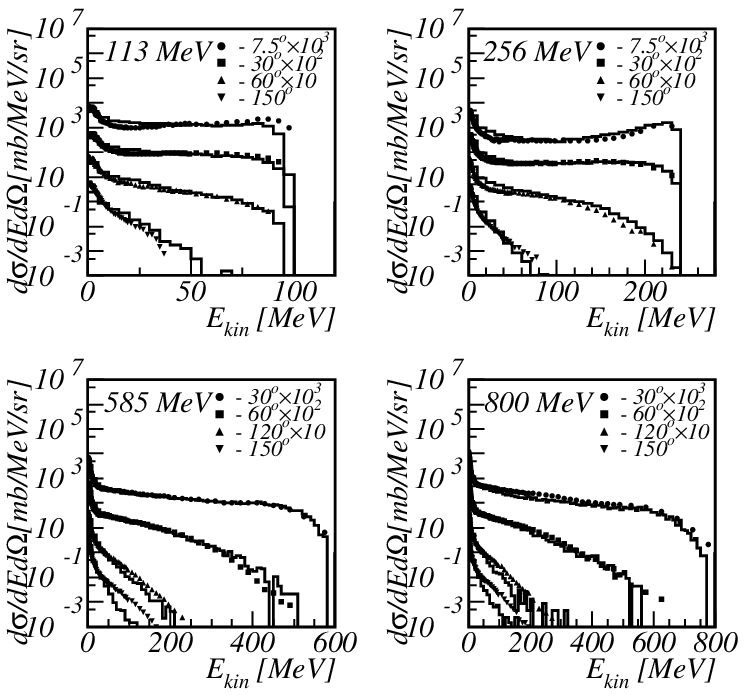}%
\caption{Double differential cross-section for neutrons
produced in proton scattering off aluminum. Histograms - Binary
Cascade predictions, points - data \cite{Meier89,Meier92,Amian93,Amian92}. \label{fig5}}
\end{figure*}

\begin{figure*}
\includegraphics[width=120mm,height=100mm]{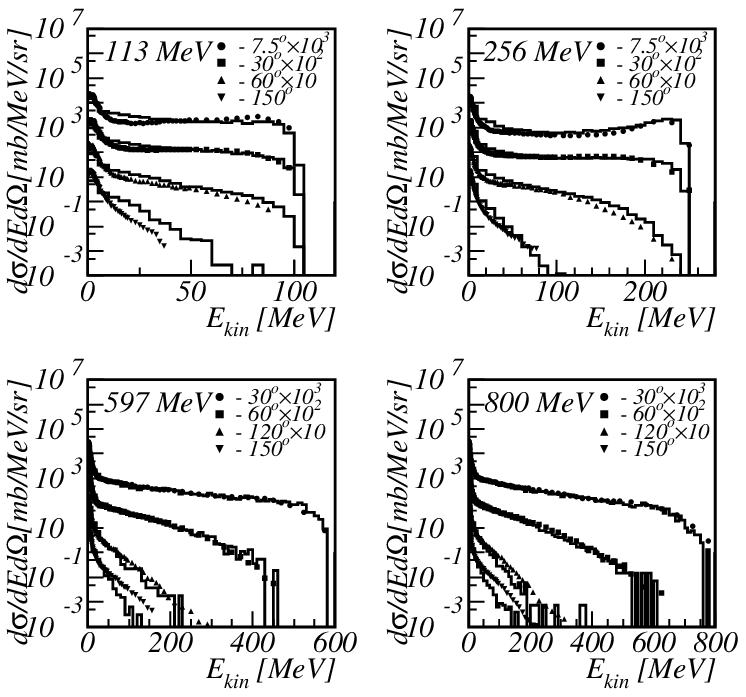}%
\caption{Double differential cross-section for neutrons
produced in proton scattering off iron. Histograms - Binary
Cascade predictions, points - data \cite{Meier89,Meier92,Amian93,Amian92}. \label{fig6}}
\end{figure*}

\begin{figure*}
\includegraphics[width=120mm,height=100mm]{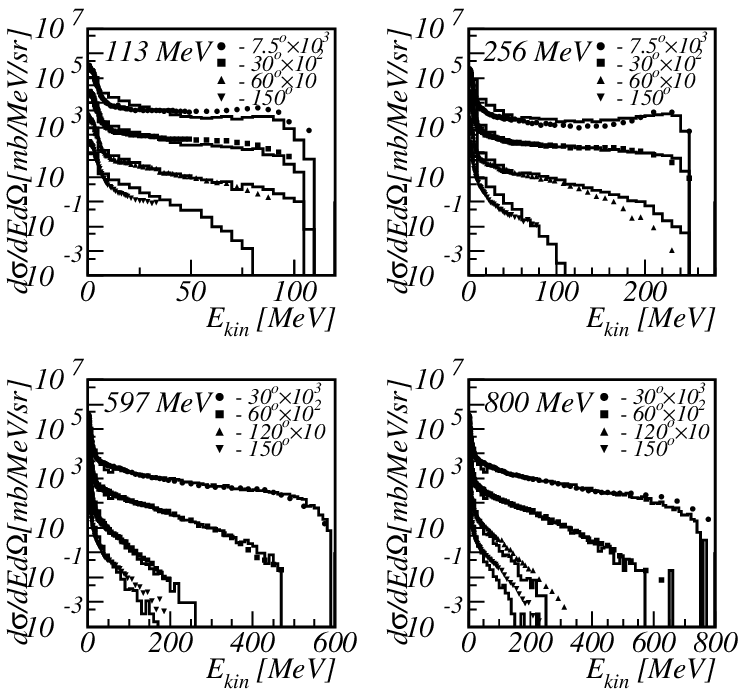}%
\caption{Double differential cross-section for neutrons
produced in proton scattering off lead. Histograms - Binary
Cascade predictions, points - data \cite{Meier89,Meier92,Amian93,Amian92}. \label{fig7}}
\end{figure*}

Additional verification test cases indicate that the Binary cascade approach is
reasonably accurate for other angles and energies as well.  Fig. \ref{fig5} shows 
comparisons for aluminum at forward and backward neutron angles at energies of 113,
256, 585 and 800 MeV.  At 113 and 256 MeV, agreement at all angles
is good.  At 585 and 800 MeV the backward angle spectra are not well-reproduced.
The same set of plots is shown in Fig. \ref{fig6} for iron, and in Fig. \ref{fig7}
for lead.  For both targets the same trends apply as were observed in aluminum.

\subsection{Pion production}

\begin{figure*}
\centering
\includegraphics[width=120mm,height=100mm]{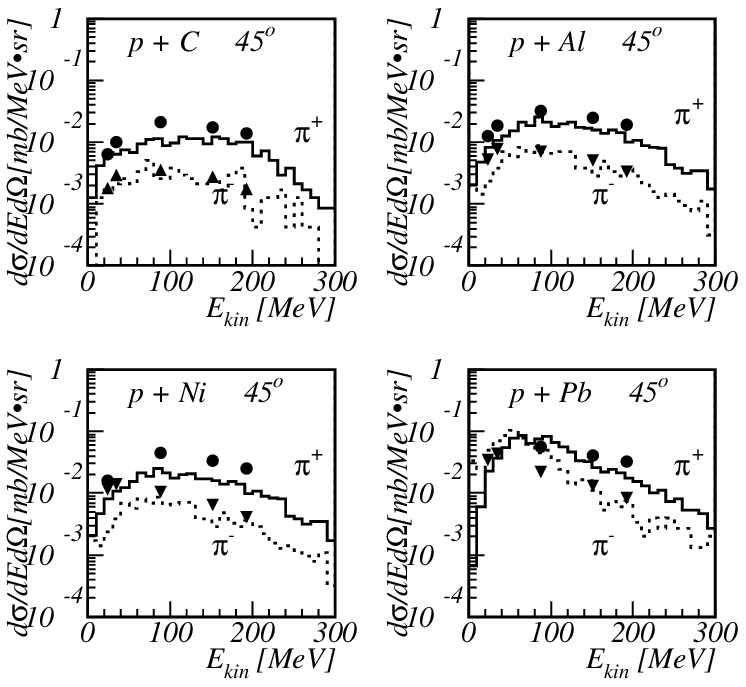}%
\caption{Double differential cross-section for pions
produced at $45^o$ in 597 MeV proton scattering off various materials.
Histograms - Binary
Cascade predictions, points - data \cite{Crawford80}. \label{fig8}}
\end{figure*}

Another useful test of the cascade codes is to look at the double differential cross
sections for the production of $\pi^+$ and $\pi^-$.  Experimental cross sections show
that $\pi^+$ production by protons is significantly larger than that for $\pi^-$.  This
feature is well-reproduced by the Binary cascade as shown in Fig.\ref{fig8}.  The
trend of the cross section versus energy is also reproduced, although the overall
normalization is underestimated by a factor of 2-3 for carbon, aluminum and nickel.

\begin{figure*}%
\centering
\includegraphics[width=120mm,height=100mm]{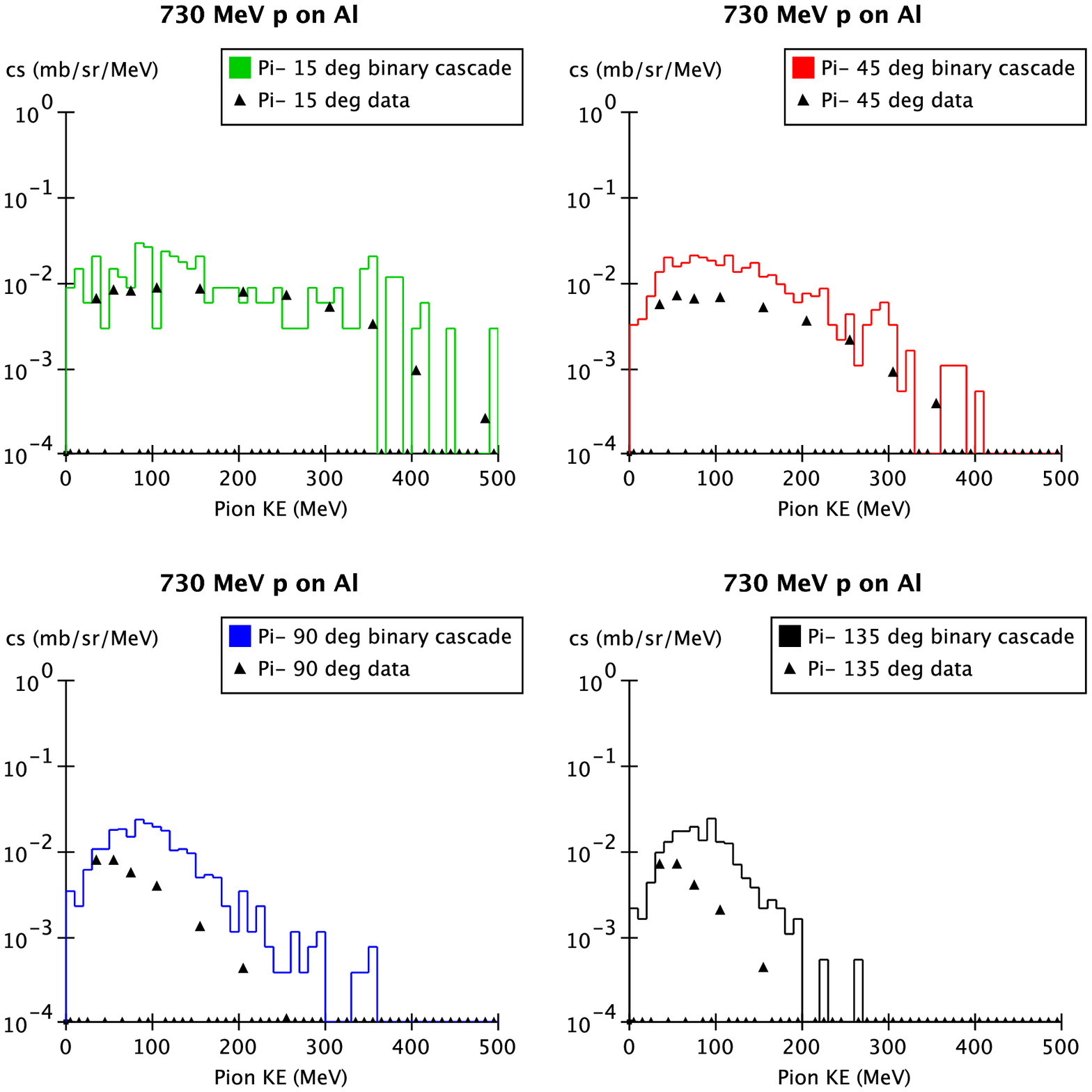}%
\caption{Double differential cross-section for $\pi^-$
produced  in 730 MeV proton scattering off aluminum.
Histograms - Binary
Cascade predictions, points - data \cite{Al730}. \label{fig9}}
\end{figure*}

\begin{figure*}
\centering
\includegraphics[width=120mm,height=100mm]{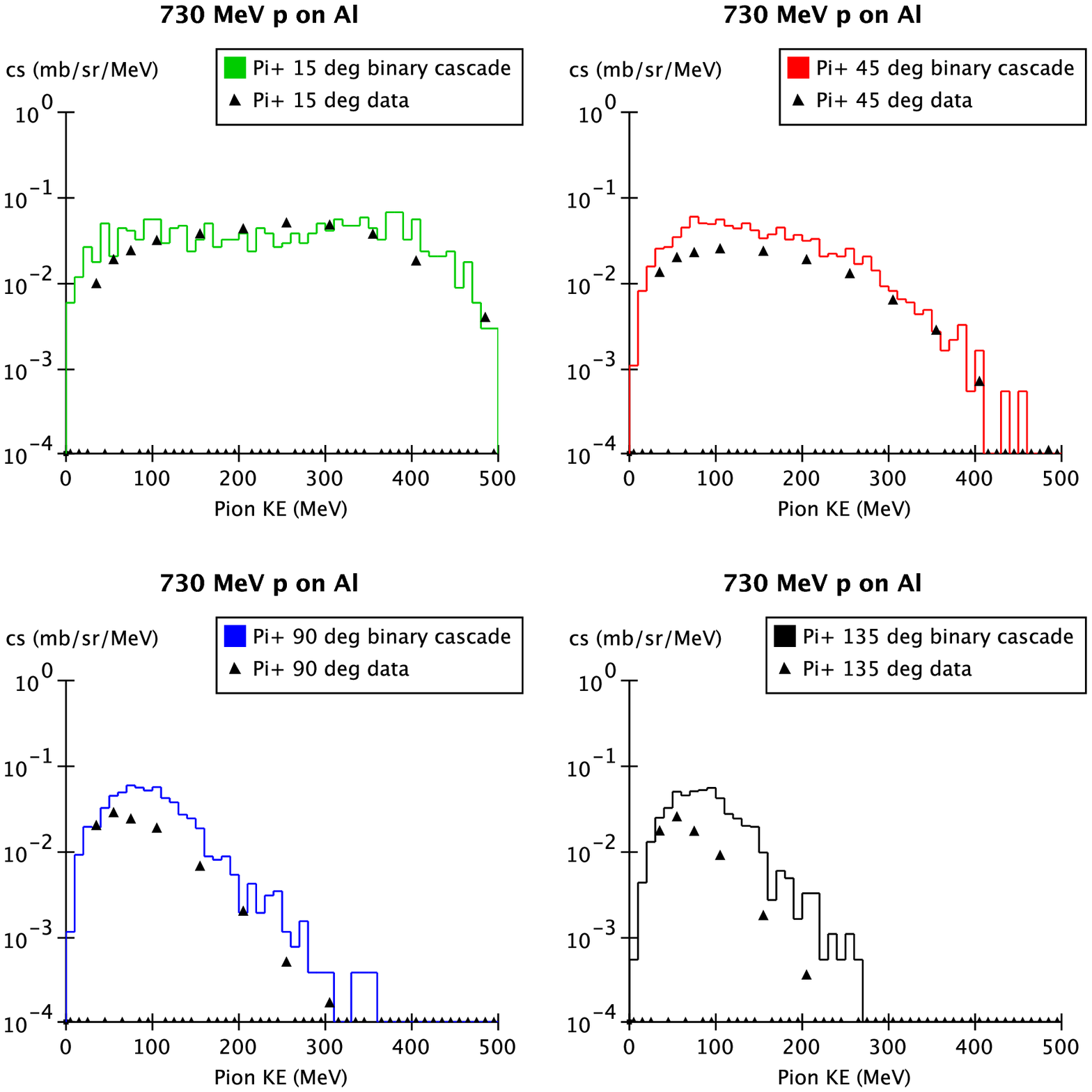}%
\caption{Double differential cross-section for $\pi^+$
produced  in 730 MeV proton scattering off aluminum.
Histograms - Binary
Cascade predictions, points - data \cite{Al730}. \label{fig10}}
\end{figure*}

\begin{figure*}
\centering
\includegraphics[width=120mm,height=100mm]{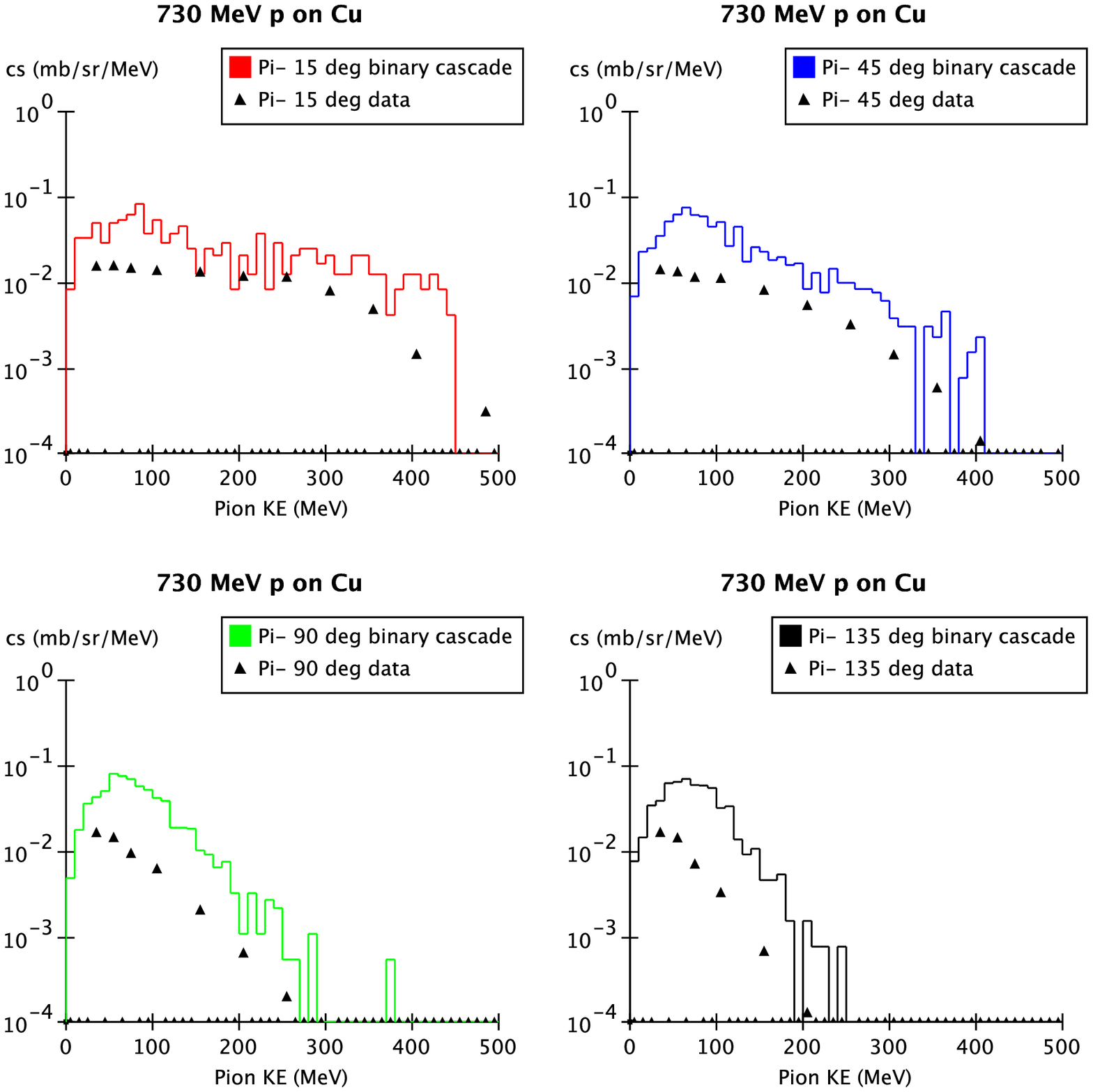}%
\caption{Double differential cross-section for $\pi^-$
produced  in 730 MeV proton scattering off copper.
Histograms - Binary
Cascade predictions, points - data \cite{Al730}. \label{fig11}}
\end{figure*}

\begin{figure*}
\centering
\includegraphics[width=120mm,height=100mm]{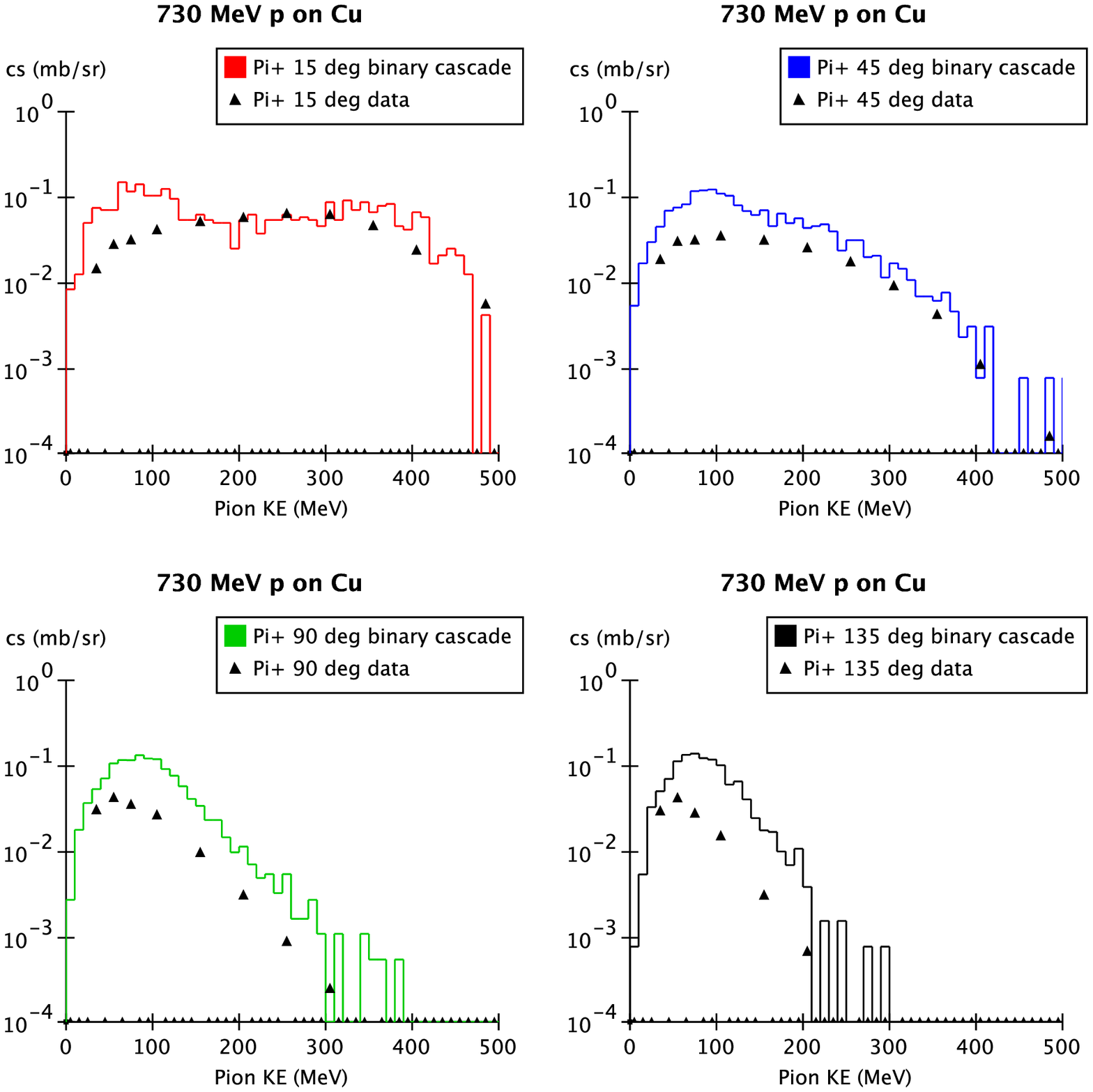}%
\caption{Double differential cross-section for $\pi^+$
produced  in 730 MeV proton scattering off copper.
Histograms - Binary
Cascade predictions, points - data \cite{Al730}. \label{fig12}}
\end{figure*}

Similar plots for $\pi^+$ and $\pi^-$ are shown for an incident proton energy of 730 MeV.
Figs.\ref{fig9} and \ref{fig10} show $\pi^-$ and $\pi^+$ at forward and backward angles from
aluminum, while Figs. \ref{fig11} and \ref{fig12} show $\pi^-$ and $\pi^+$ from copper.  In
all these cases the observed ratio of $\pi^+$ to $\pi^-$ production is reproduced.  As was
the case for 597 MeV incident protons (Fig. \ref{fig8}), the trend of cross section versus
energy is also reproduced.  However at 730 MeV it is seen that the Binary cascade does an
increasingly poor job of reproducing the overall normalization as the pion angle increases.

\begin{figure*}
\centering
\includegraphics[width=120mm,height=100mm]{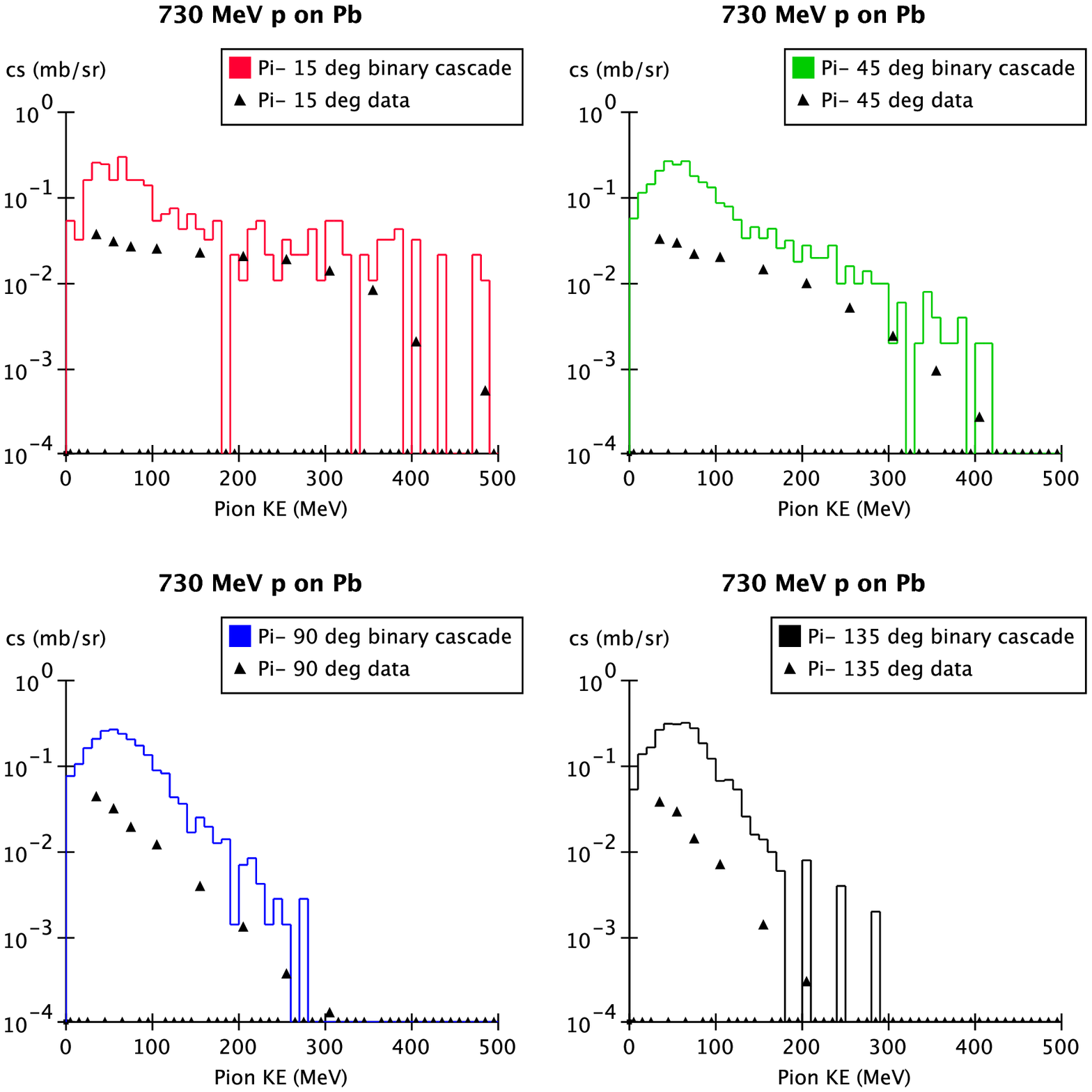}%
\caption{Double differential cross-section for $\pi^-$
produced  in 730 MeV proton scattering off aluminum.
Histograms - Binary
Cascade predictions, points - data \cite{Al730}. \label{fig13}}
\end{figure*}

\begin{figure*}
\centering
\includegraphics[width=120mm,height=100mm]{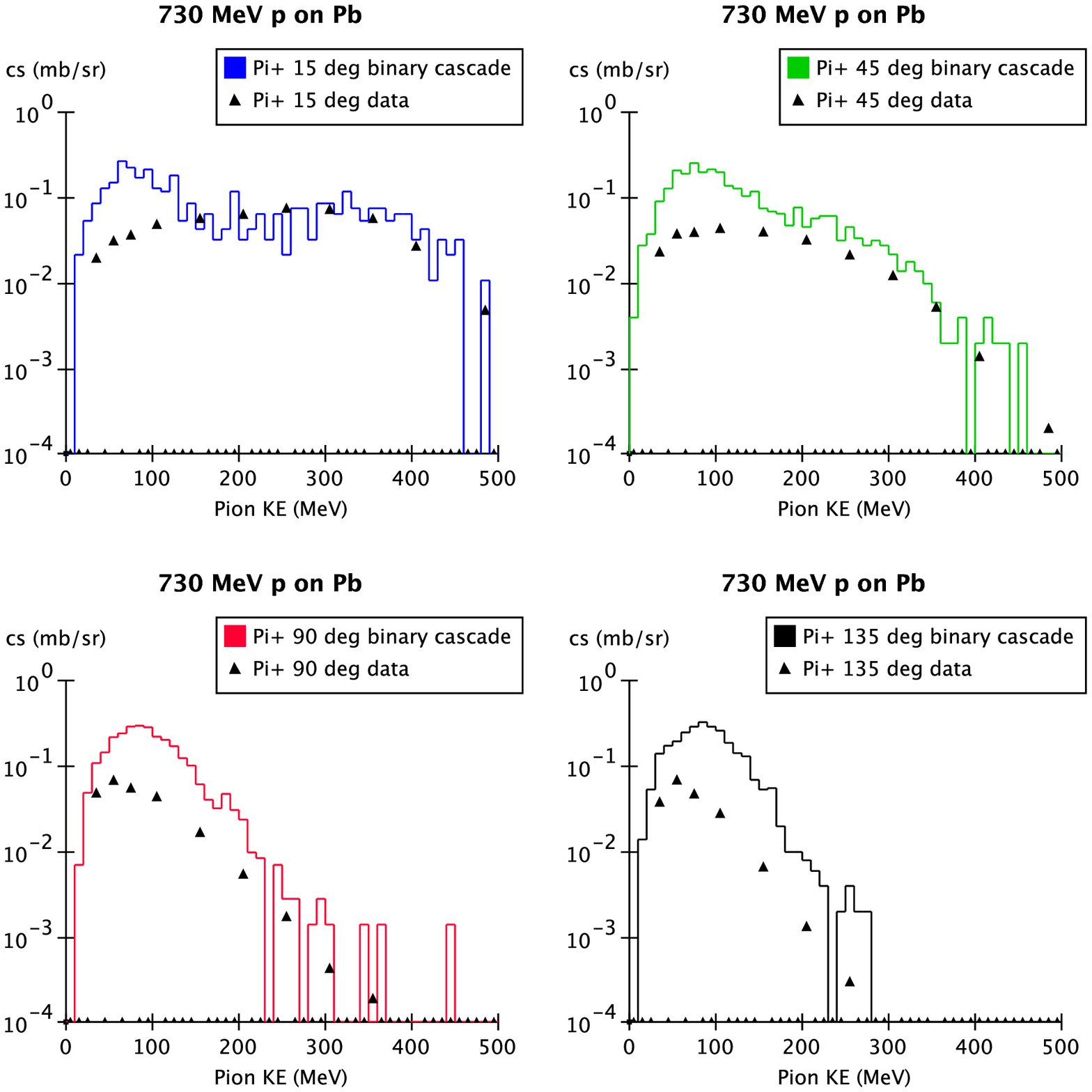}%
\caption{Double differential cross-section for $\pi^+$
produced  in 730 MeV proton scattering off aluminum.
Histograms - Binary
Cascade predictions, points - data \cite{Al730}. \label{fig14}}
\end{figure*}

\subsection{Future verification work}

While the existing verification suite has already been used extensively,
further development is required.  So far the validity of the cascade codes
has only been demonstrated for incident protons at the low end of their
energy ranges.  In order to test the full range of energies and particles,
many more test cases are required, including:
\begin{itemize}
\item incident proton energies up to 15 GeV,
\item incident pion energies up to 15 GeV.
\item incident neutrons, and
\item proton inclusive spectra.
\end{itemize}

\section{CONCLUSIONS}

A verification suite for hadronic interaction models in the cascade energy region has
been designed and implemented.  This suite has been used both to develop the Bertini and Binary
cascade model codes for the Geant4 toolkit and to compare their predictions with data from thin
target scattering experiments.

The suite has proved to be an efficient tool for model validation which can be used for the
development and testing of other hadronic interaction models.
The suite is also being expanded to include tests for more types of incident particles and
higher energies.
 \\

\begin{acknowledgments}

Work of VI was partially supported by INTAS (grant INTAS-2001-0323).

The work of TK and DHW was supported by U.S. Department of
Energy contract DE-AC03-76SF00515.
  \\

\end{acknowledgments}

\end{document}